\def\year{2019}\relax
\documentclass[letterpaper]{article}
\usepackage{aaai19}
\usepackage{times}
\usepackage{helvet}
\usepackage{courier}
\usepackage{url}
\usepackage{graphicx}
\usepackage{booktabs} 
\usepackage{url}
\usepackage{subcaption}
\usepackage{latexsym}
\usepackage{graphicx}
\usepackage{array}
\usepackage[font=small]{caption}
\usepackage{microtype}
\usepackage{multirow}
\usepackage{amsfonts}
\usepackage{amsmath}
\usepackage{amssymb}
\usepackage{enumerate}
\usepackage{enumitem}
\usepackage{xcolor}
\usepackage{color, colortbl}
\definecolor{LGray}{gray}{0.9}
\definecolor{Gray}{gray}{0.8}
\definecolor{DGray}{gray}{0.7}

\usepackage{titlesec}
\titlespacing{\paragraph}{%
  0em}{
  0.1\baselineskip}{
  0.2\baselineskip}%

\title{What Twitter Profile and Posted Images Reveal \\ About Depression and Anxiety}
\author{Sharath Chandra Guntuku$^{1}$, Daniel Preotiuc-Pietro$^{2}$, Johannes C. Eichstaedt$^{1}$, Lyle H. Ungar$^{1}$\\
       $^{1}$University of Pennsylvania, $^{2}$Bloomberg LP\\
       \{sharathg@sas, jeich@sas, ungar@cis\}.upenn.edu, dpreotiucpie@bloomberg.net
       }
\begin{document}
\maketitle

\begin{abstract}
Previous work has found strong links between the choice of social media images and users' emotions, demographics and personality traits. In this study, we examine which attributes of profile and posted images are associated with depression and anxiety of Twitter users. 
We used a sample of 28,749 Facebook users to build a language prediction model of survey-reported depression and anxiety, and validated it on Twitter on a sample of 887 users who had taken anxiety and depression surveys. 
We then applied it to a different set of 4,132 Twitter users to impute language-based depression and anxiety labels, and extracted interpretable features of posted and profile pictures to uncover the associations with users' depression and anxiety, controlling for demographics. 
For depression, we find that profile pictures suppress positive emotions rather than display more negative emotions, likely because of social media self-presentation biases. They also tend to show the single face of the user (rather than show her in groups of friends), marking increased focus on the self, emblematic for depression. Posted images are dominated by grayscale and low aesthetic cohesion across a variety of image features. 
Profile images of anxious users are similarly marked by grayscale and low aesthetic cohesion, but less so than those of depressed users. 
Finally, we show that image features can be used to predict depression and anxiety, and that multitask learning that includes a joint modeling of demographics improves prediction performance.
Overall, we find that the image attributes that mark depression and anxiety offer a rich lens into these conditions largely congruent with the psychological literature, and that images on Twitter allow inferences about the mental health status of users.
\end{abstract}

\maketitle

\section{Introduction}

Depression continues to be under-diagnosed with only 13-49\% receiving minimally adequate treatment  \cite{wang2005twelve}. This is caused by a complex set of factors such as imperfect screening methods, social stigma associated with the diagnosis, and lack of cost-effective resources for and access to diagnosis and treatment. Efforts to detect depression predominantly rely on online and phone surveys. However, these surveys are resource intensive, both in terms of cost and time, and insufficiently reach all at risk  \cite{suchman1962analysis}. The single time point of assessment also means there are often gaps between data collection and depression onset. Automated analysis of user generated content can potentially provide methods for early detection of depression \cite{guntuku2017detecting}. If an automated process could detect elevated depression or anxiety levels in a person, that individual could be targeted for a more thorough assessment (and provided with digitized forms of support and treatment), alleviating many of the constraints associated with traditional assessment methods.

\begin{figure}[!t]
  \centering
  \includegraphics[width=.9\columnwidth]{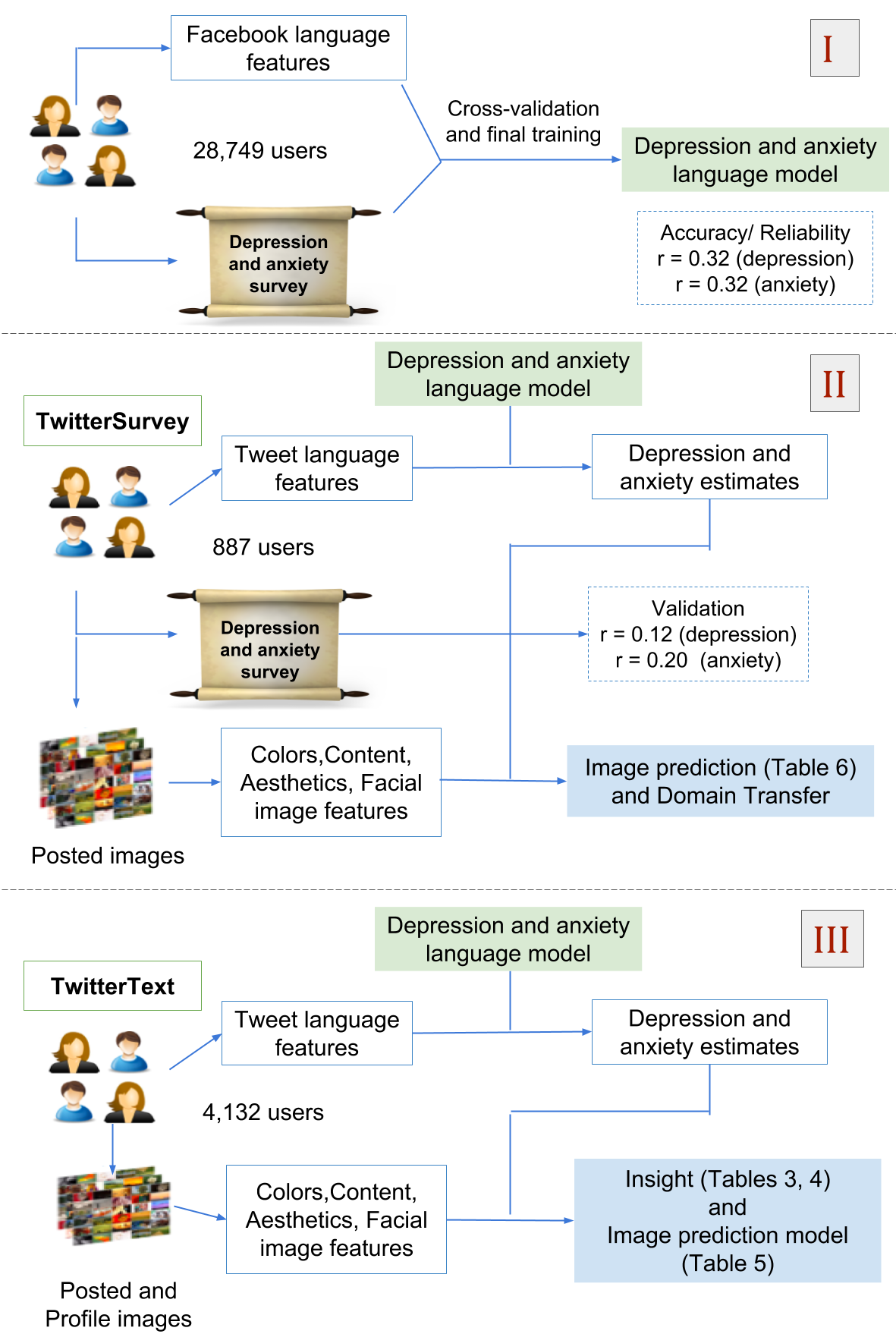}
\caption{Overview of data flow, models and analysis in the paper}
\label{fig:overview}
\end{figure}

\begin{table*}[!t]
\centering
\resizebox{\textwidth}{!}{
\small
\begin{tabular}{|l|c|c|c|c|}
\hline
\rowcolor{Gray} \textbf{Study} & \textbf{\# Users} & \textbf{Traits} & \textbf{Image Type} & \textbf{Image Features}\\
Our Work & \textbf{887 + 4,132} & Depression \& Anxiety & Twitter Posted \& Profile Images & Color, Facial, Aesthetics, Content, VGG-Net \\
\cite{reece2017instagram} & 166 & Depression & Instagram Photos & Colors \\
\cite{andalibi2015depression} & -- & Depression & 500 `\#depression' Instagram Photos & Manual Annotation \\  \hline
\cite{ferwerda2018you} & 193 & Personality (continuous) & Instagram Photos &  Content \\
\cite{nie2018understanding} & 2238 & Perceived Personality (continuous) & Web Portrait Images & Facial, Social information \\
\cite{samani2018cross} & 300 & Personality (continuous) & Twitter and Flickr Posts, Likes, \& Profiles & Colors, Content, VGG-Net \\
\cite{farnadi2018user} & 5670 & Personality (binary) & Facebook Profile Images & Facial, Text, Likes \\
\cite{guntuku2017studying} & 4132 + 161 & Personality (continuous) & Posted, liked images and text on Twitter & Color, Bag of Imagga tags, VGG-Net \\
\cite{segalin2017your} & 11,736 & Personality (continuous \& binary) & Facebook Profile Images & Aesthetics, BOVW, VGG-Net, IATO  \\
\cite{persimages16icwsm} & 66,502 & Personality (continuous) & Twitter Profile Images & Color, Facial \\
\cite{ferwerda2016using} & 113 & Personality (binary)  & Instagram Photos & Colors, \#Faces, Filters \\
\cite{skowron2016fusing} & 62 & Personality (binary) & Instagram Photos &  Colors \\
\cite{guntuku2016likes} &  300 &	Personality (continuous)  &	Liked (`Fave') images on Flickr &	Colors, semantic features, aesthetics \\
\cite{guntuku2015others}&  123 &	Personality (continuous)  &	Selfies on Weibo &	Color, Aesthetics, BOVW, Emotions \\
\cite{AlMoubayed14} & 829 & Personality (binary) & Face Images & Eigenfaces \\
\cite{celli2014automatic} & 112 & Personality (binary) & Facebook Profile Images & Bag-of-Visual-Words (BOVW) \\
\hline
\end{tabular}
}
\caption{Summary of data and methods used in previous work analyzing images of individual users.}
\label{tbl:rw}
\end{table*}

The last few years have seen an increased interest in studying mental health through social media. Recent studies have investigated the association of language and social media use patterns with several mental illnesses, including stress, depression and suicidality. Such studies can be grouped into two categories: a) studies aimed at detecting mental illnesses and their symptoms from social media (the focus of this paper); and b) studies that examine how the use of social media contributes to or alleviates distress in users. Most prior works explored the use of linguistic attributes to predict mental health conditions with reasonable success.  \cite{guntuku2017detecting} review the use of social media text-based features in detecting mental health conditions. 

With the ubiquity of camera-enabled mobile devices and growth of image-based social media platforms (such as Twitter, Instagram or Snapchat) social media content is increasingly in image form  \cite{burdick2012digital_humanities}, and preliminary work suggests that users are increasingly communicating their emotions using images \cite{reece2017instagram}. Current computer vision algorithms allow us to automatically analyze large scale data sets and extract shallow color features as well as semantic content such as objects or scenes. Previous work has studied the variation in colors and use of filters of posted images in depressed individuals and has not studied  the diverse content associated with posted images or the characteristics of profile images for users with depression and anxiety  \cite{reece2017instagram}.

Moreover, prior works typically study mental health conditions in isolation, not exploiting the underlying implicit correlations in conditions and demographic factors. Tasks with underlying commonalities have been shown to benefit from multi-task learning, e.g. action recognition  \cite{wang2016human}, chronic disease prediction  \cite{nie2015beyond}. Particularly user demographics (age and gender) and the co-morbidity between mental health conditions have been used to diagnose patients by clinicians  \cite{bijl1998prevalence}. We use multi-task learning to take these strong associations into account.

In summary, we examine the following questions: 
\begin{itemize}[noitemsep,topsep=0pt,leftmargin=1em]
    \item What meaningful, relevant and interpretable patterns do images that users post and choose as profile pictures reveal about users' depression and anxiety?
    \item What effect does joint learning of demographics along with co-morbid mental health conditions have on prediction performance?
    \item Do text-predicted labels help when predicting survey-based ground-truth labels from images?
\end{itemize}

\section{Related Work}

Researchers have used images to study personality, measured using the Big Five model \cite{bigfive}, based on profile pictures (with facial features). Others have also used posted images. One of the earliest works predicted self-assessed personalities of 100 users using their Facebook profile images  \cite{celli2014automatic} with $\sim$65\% accuracy using bag-of-visual-words (BoVW) features. Random portraits from the web \cite{nie2014your} and existing face recognition data sets \cite{AlMoubayed14} were also used to model users' personality. Recently, aesthetic features \cite{datta2006,Machajdik2010} (apart from facial features) were also used to predict personality on a $\sim$66,000 user data set \cite{persimages16icwsm} from Twitter. Further,  \cite{segalin2017your} used multiple sets of features extracted from Facebook images of 11,736 users and built computational models which have better predictive power than human raters in predicting personality traits, specifically Extraversion and Neuroticism. 

However, in the health domain manifestation of mental
health conditions in individual users based on social media images is under explored, despite recent work being done on studying public health of communities  \cite{Manikonda2017,chancellor2017multimodal,garimella2016social}. Table \ref{tbl:rw} presents a summary of the relevant works: number of users, traits, image types studied and features used. 

\cite{andalibi2015depression} examined the distribution of themes in 500 images with the hash-tag \#depression uncovering themes posted by depressed users. However, limited to a small set of images, further work is required to study the generalisability of the findings. \cite{reece2017instagram} looked for the markers of depression in Instagram posts. However, the study aggregates images by days of posting instead of participants even though depression was measured a single time as a trait using a questionnaire. It also looks at a limited set of image features on posted images.

The aim of our work is to use social media images to study how depression and anxiety are related to the content of images that people post or choose as profile pictures on social media. We explore if incorporating author attributes such as age and gender in the joint modeling of depression and anxiety can improve prediction performance. Further, we examine if weak text-predicted labels can improve performance in predicting reliable survey based outcomes. 

\section{Data}

We use a Facebook dataset (where we have language data but not image data from users) and two Twitter datasets (where we have both language and image data from users) in this study (overview in Figure \ref{fig:overview}). This study received approval from the University of Pennsylvania Institutional Review Board (IRB). 

The Facebook data is from a previous study \cite{schwartz2014towards}, consisting of 28,749 users who had taken the IPIP NEO-PI-R survey \cite{neopir} that contains the Depression and Anxiety Facets of the Neuroticism Factor. These users also consented to share access to their status updates which was used to build a text-regression model. The text model was trained using 1-,2-, and 3-grams used by at least 5\% of users (resulting in 10,450 ngrams), 2000 LDA derived topics, and 64 LIWC features extracted from status updates and a linear regression with L2 ridge penalization on 10\% principal component features.  In the original validation, the model achieved a Pearson correlation of $r = .32$ predictive performance, which is considered a high correlation in psychology, especially when measuring internal states \cite{psychassessment}. We did not have access to image data for the Facebook cohort. 

TwitterSurvey was collected by deploying a survey on Qualtrics\footnote{www.qualtrics.com/Survey-Software} (a platform similar to Amazon Mechanical Turk), comprising several demographic questions (age, gender, race, education, and income) and the Beck's Depression Inventory  \cite{beck1996beck} to measure depression and Anxiety facet from the International Personality Item Pool proxy for the NEO Personality Inventory Revised (NEO-PI-R) \cite{neopir}. Users received an incentive for their participation, and we obtained their informed consent to access their Twitter posts. All users were based in the US. Out of 887 users who took the survey, 766 users posted 241,683 images. We excluded users who posted fewer than 20 images, resulting in 560 users with 145,436 posted images. We used continuous depression and anxiety scores (descriptive statistics in Table \ref{tab:desc}) and, for regression, we standardized the scores by converting them to have mean 0 and standard deviation 1. As the distribution of all psychological traits is approximately normal with basically no exceptions, we preferred not to threshold continuous distributions, to have the largest amount of useful variance available when considering covariances and correlations. In Figure \ref{fig:dep_post_eg}, we visualize random images from users with top and bottom 25\% percentile of both depression and anxiety scores. The distribution of scores are shown in Table \ref{tab:desc}.

TwitterText is a domain related dataset which has self-reported age and gender for all users. We used it to be able to control for and use demographics for multi-task modeling of mental health conditions. This dataset was used in previous works by \cite{preoctiuc2017beyond,guntuku2017studying}. It is an order of magnitude larger data set consisting of 4132 Twitter users. Since we do not have depression and anxiety computed via surveys for this data set, we use the language-prediction model from Facebook to impute depression and anxiety scores. We downloaded the 3200 most recent user tweets for each user, leading to a data set of 5,547,510 tweets, out of which 700,630 posts contained images and 1 profile image each across 3498 users. We exclude the texts associated with the tweets which contain images when predicting depression and anxiety to limit any potential confound. Then, for our analysis, we excluded users who posted less than 20 photos. 

\begin{table}[t!]
\centering
\small
\resizebox{\columnwidth}{!}{
\begin{tabular}{|c|l|l|l|l|l|}
\hline
\multicolumn{3}{|c|}{\cellcolor[HTML]{C0C0C0}{\color[HTML]{000000} \textbf{TwitterSurvey}}}                         & \multicolumn{3}{c|}{\cellcolor[HTML]{C0C0C0}{\color[HTML]{000000} \textbf{TwitterText}}}                           \\ \hline
\multicolumn{1}{|c|}{\textbf{}} & \multicolumn{1}{c|}{\textbf{Dep Score}} & \multicolumn{1}{c|}{\textbf{Anx Score}} & \multicolumn{1}{c|}{\textbf{}} & \multicolumn{1}{c|}{\textbf{Dep Score}} & \multicolumn{1}{c|}{\textbf{Anx Score}} \\ \hline
\textbf{Min}                    & -1.323                                  & -2.390                                  & \textbf{Min}                   & -2.644                                  & -3.052                                  \\ \hline
\textbf{25\%}                   & -0.758                                  & -0.693                                  & \textbf{25\%}                  & -0.719                                  & -0.729                                  \\ \hline
\textbf{50\%}                   & -0.194                                  & 0.155                                   & \textbf{50\%}                  & -0.084                                  & -0.077                                  \\ \hline
\textbf{75\%}                   & 0.629                                   & 0.791                                   & \textbf{75\%}                  & 0.702                                   & 0.662                                   \\ \hline
\textbf{Max}                    & 3.851                                   & 1.852                                   & \textbf{Max}                   & 3.492                                   & 3.967                                   \\ \hline
\end{tabular}
}
\caption{\centering Descriptive statistics of outcomes in both data sets. Scores are z-normalized}
\label{tab:desc}
\end{table}

\section{Feature Extraction}
Since we are dealing with both posted images and profile pictures, we extract different sets of features to capture the representations associated with both. From posted images, we extract colors, aesthetics and image content related features, and from profile images, in addition to the three sets, we also extract facial-related features considering the literature linking facial expressions to emotions and mental states \cite{gur1992facial}. 

\subsection{Colors}

The colors of an image represent the most notable features to a human. Research has shown that colors can invoke emotions \cite{wexner1954degree}, psychological traits \cite{huang2006} or, on social media, reflect the personality of the person posting the image \cite{skowron2016fusing} and even mental health states \cite{reece2017instagram}. 

Colors can be expressed in various color spaces. We use the HSV (Hue--Saturation--Value), which provides a more intuitive representation of colors for humans \cite{bigun2006vision}. A pixel in the HSV space is characterized by three numbers: (1) \textit{Hue}: the color type ranging between 0 and 360 degrees e.g.,\ 0 is red, 60 is yellow; (2) \textit{Saturation}: the intensity of the color ranging from 0 to 1 e.g.,\ 0 represents no color and is a shade of gray; (3) \textit{Value}: the brightness of the color ranging from 0 to 1 e.g.,\ 0 represents black.

Using the images converted to the HSV space, we first determine whether an image is \textit{grayscale}. A picture is considered grayscale if there is no pixel for which hue can be calculated with accuracy i.e.,\ $ V \in [0.15, 0.95], S > 0.2$ \cite{ke2006design}. We ignore grayscale images ($216$ profile images, 4.99\% of posted images) from subsequent color analysis only, as their presence may bias the results.
We compute \textit{saturation} and \textit{brightness} as the average saturation and value respectively of the pixels in the image. An experimental study of colors established the relationship between saturation and brightness and the dimensional model of affect containing three factors: \textit{Pleasure} $= .69 \cdot V + .22 \cdot S$, \textit{Arousal} $= -.31 \cdot V + .60 \cdot S$, \textit{Dominance} $= -.76 \cdot V + .32 \cdot S$ \cite{mehrabian1974approach,valdez1994effects}. We compute the standard deviation of the HSV values, which we will use only for prediction.

Using the hues of an image, we compute the \textit{hue count} of a photo as a measure of its simplicity \cite{ke2006design}. Professional photos usually have a lower number of unique hues, although each color may be rich in tones, leading to a simpler composition. To compute the hue count defined in \cite{ke2006design}, we obtain the 20-bin hue histogram from each pixel for which we can accurately compute its hue (see above) and compute the number of bins containing at least 5\% of the pixels of the maximum bin size. As the hue count distribution is skewed towards a low number of hues, we log-scale this value for our analysis.

We also compute a 6-bin histogram which splits pixels into the primary and secondary colors
and a 12-bin histogram to count the proportion of pixels from the primary, secondary and tertiary colors. Finally, we compute percentage of warm and cold color pixels using the hue values: \textit{Warm}: $H \in [285,75]$, \textit{Cold}: $H \in [105,255]$. A combination of contrast and hue count was used as a proxy for sharpness. After extracting features from individual posted images, we aggregate them to the users using mean pooling.

\subsection{Aesthetics}

Using pair-wise ranking of image pairs as well as
the image attribute and content information,  \cite{kong2016aesthetics} propose to learn aesthetics. Two base networks of the Siamese architecture for each of the two images (in the pair) adopt the AlexNet configurations with the final fully connected layer removed.
The base network is fine-tuned using aesthetics data with an Euclidean Loss regression layer followed by the Siamese network ranking the loss for every sampled image pairs. The fine-tuned network is used as a preliminary feature extractor. These features are then used for an attribute prediction task which is trained in a multi-task manner by combining the rating regression Euclidean loss, attribute classification loss and ranking loss. Finally, a content classification softmax layer is added to predict a predefined set of category labels. The categories are defined as: `balancing element' - whether the image contains balanced elements; `content' - whether the image has good/interesting content; `color harmony' - whether the overall color of the image is harmonious; `object emphasis' - whether the image emphasizes foreground objects; `rule of thirds' - whether the photography follows rule of thirds; `vivid color' - whether the photo has vivid color, not necessarily harmonious color; `repetition' - whether the image has repetitive patterns; `symmetry' - whether the photo has symmetric patterns; `depth of field' - whether the image has shallow depth of field; `lighting' - whether the image has good/interesting lighting; `motion blur' - whether the image has motion blur. The various categories labeled are photographic attributes and image content information which help regularize the photo aesthetic score, which is modeled as a complex non-linear combination of each of the categories. The network also outputs an aesthetic score. After extracting features from individual posted images, we aggregate them to the users using mean pooling.

\begin{figure*}[!t]
	\begin{subfigure}{.5\textwidth}
		\centering
		\includegraphics[width=.9\linewidth]{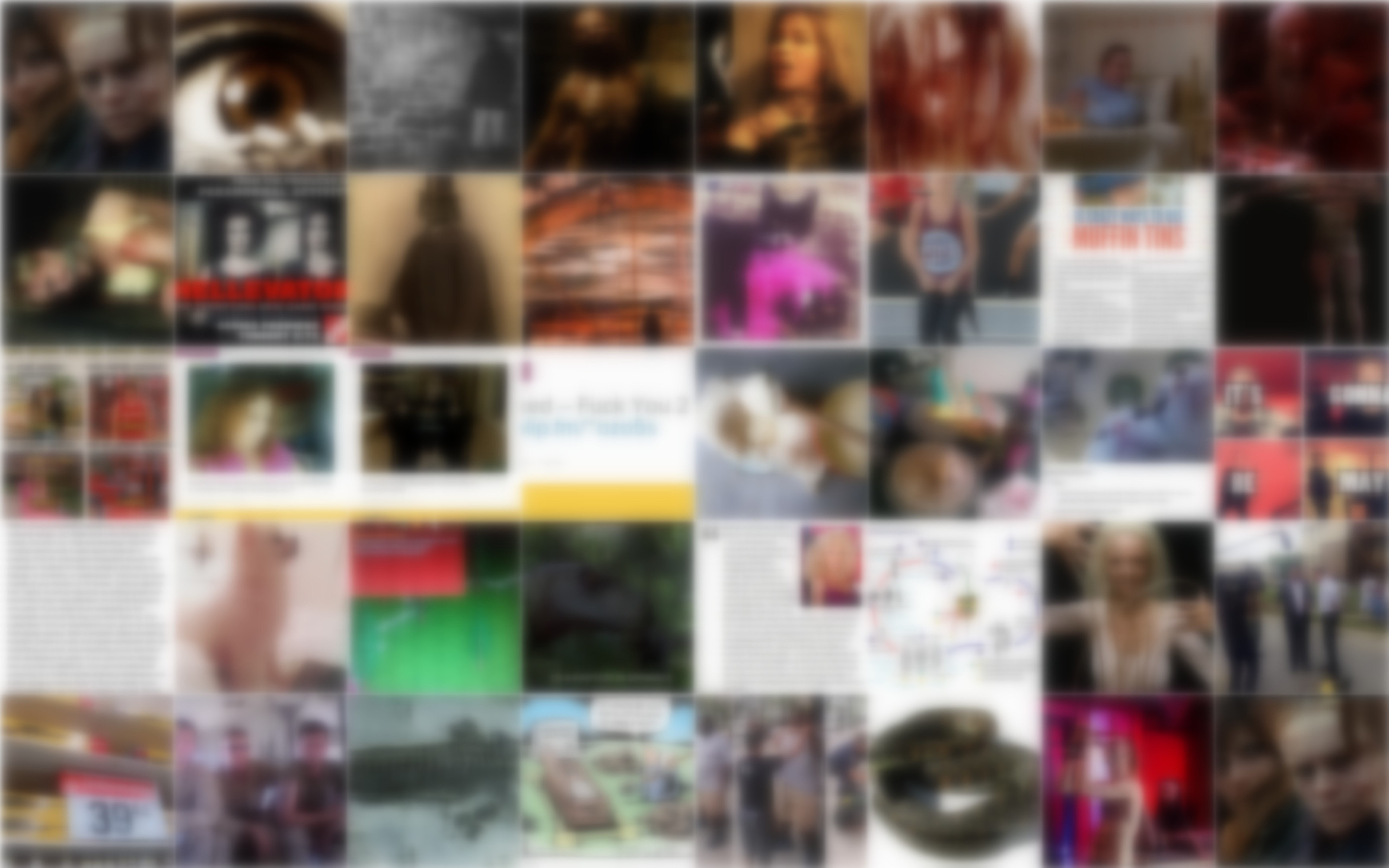}
		\caption{High level of depression and anxiety}
		\label{fig:high_dep_eg}
	\end{subfigure}%
	\begin{subfigure}{.5\textwidth}
		\centering
		\includegraphics[width=.9\linewidth]{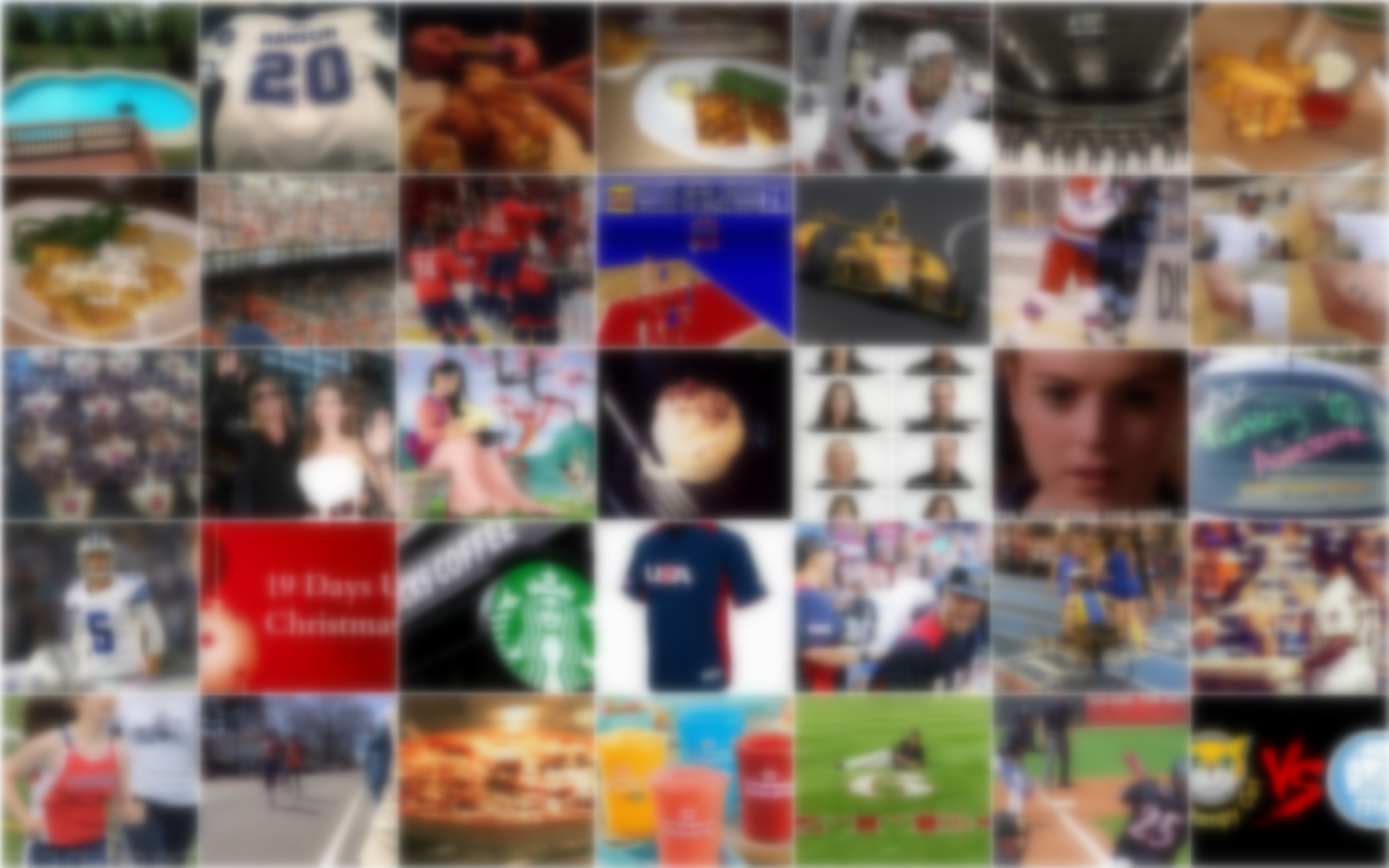}
		\caption{Low level of depression and anxiety}
		\label{fig:low_dep_eg}
	\end{subfigure}
	\caption{A random sampling of images posted by users with top and bottom 25\% percentile of both depression and anxiety scores, blurred for privacy reasons.}
	\label{fig:dep_post_eg}
\end{figure*}

\subsection{Image Content} 

Images posted in tweets can have diverse content beyond faces. Therefore, apart from color features, as used in prior works \cite{reece2017instagram} and \cite{ferwerda2016using}), we use automatic content analysis techniques to generate tags for these images. We labeled all images with the Imagga Tagging API\footnote{\url{http://docs.imagga.com/\#auto-tagging}} and generated for each image a \textit{bag-of-tags} out of the top-10 predicted tags, following the developers' recommendations. We removed all tags that occurred less than 200 times in our data set, leaving us with 1,299 distinct tags. Imagga was successfully used in previous research \cite{garimella2016}.

We found multiple tags which were very similar and usually co-occurred together (e.g.,\ glass, drink, beverage; waves, shoreline, seascape, seaside). We therefore reduce the feature space by learning tag clusters that contain frequently co-occurring tags. This decreases sparsity and increases interpretability. We use a clustering procedure that was originally applied to words in tweets, which produced very accurate prediction results \cite{impact14eacl}. First, we compute the Normalised Pointwise Mutual Information (NPMI) between all pairs of tags \cite{bouma2009npmi}. NPMI measures the degree to which two tags are likely co-occur in the same context (image) and takes a maximum value of $1$ if two tags always co-occur and a value of $0$ if they occur as according to chance. We use NPMI as a similarity measure; negative values are replaced with 0 as in \cite{impact14eacl} and compute a tag $\times$ tag similarity matrix. We feed this matrix to the spectral clustering algorithm, a hard-clustering method appropriate for generating non-convex clusters \cite{Ng02,Shi00,VonLuxburg07}, which performs a graph partitioning on the Laplacian of the similarity matrix. The number of clusters needs to be specified in advance. We use 400 clusters throughout this study, based on preliminary experiments.

Once these clusters of semantically similar tags are created, we represent each image as a vector containing the normalized number of times each tag cluster is detected. For each user, we derive a feature vector of \textit{image content topics} as the normalized number of times each topic cluster is present in the tweet-embedded images. We also calculated the percentage of image posts and percentage of posts with people (based on Imagga tags). 

\paragraph{VGG-Net} We use a pre-trained version of the 19-layer version of the VGG-Net image classifier based on convolutional neural networks \cite{simonyan14c}. This classifier achieved the best results in the ImageNet Large Scale Visual Recognition Challenge 2014 in the object classification and localization challenge. It predicts a class probability for the 1,000 objects in the ImageNet tagset. We also extract the features from the last fully connected layer (\textit{fc7}) for prediction. After extracting features from individual posted images, we aggregate them to the users using mean pooling.

\subsection{Face-related}

For profile photo analysis, we use the Face++\footnote{\url{http://faceplusplus.com/}} and EmoVu\footnote{\url{http://emovu.com/}} APIs for facial feature extraction, as profile photos on Twitter usually contain one or more faces. Previous research has shown facial features can predict personality of users \cite{persimages16icwsm} and has used Face++ as a proxy for extracting user demographics from profile images  \cite{an2016greysanatomy,zagheni14}.

\paragraph{Facial Presentation \& Expression} This category contains facial features that try to capture the self-presentation characteristics of the user. Features include the face ratio (the size of the face divided by the size of the image), whether the face features any type of glasses (reading or sunglasses), the closeness of the subject's face from the acquisition sensor provided by EmoVU's attention measurement, the eye openness and the 3D face posture, which includes the pitch, roll and yaw angle of the face.

For facial emotion detection, we use Ekman's model of six discrete basic emotions: anger, disgust, fear, joy, sadness and surprise, which were originally identified based on facial expressions \cite{ekman1971constants}. We use the EmoVU API to automatically extract these emotions, as well as neutral expression \cite{batty2003early}, from the largest detected face in each profile image. The six basic emotions can be categorized as either positive (joy and surprise) or negative (anger, disgust, fear, sadness). Along with the basic emotional expressions, EmoVU also gives composite features: Expressiveness is the highest value of the six basic emotions; Negative and positive mood are calculated as the maximum value of the positive and negative emotions respectively; valence is the average of the negative mood and positive mood. Also, we add the smiling degree provided by Face++.

\begin{table}[htp!]
\centering
\small
\begin{subtable}{\columnwidth}
\begin{tabular}{|l!{\vrule width 1pt}c|c!{\vrule width 1pt}c|c|}
\hline
\rowcolor{DGray}
\textbf{Feature} & \multicolumn{2}{c!{\vrule width 1pt}}{\textbf{Demographics}} & \multicolumn{2}{c|}{\textbf{MentalHealth}}\\
\hline
\rowcolor{Gray}
\textbf{Colors}  & Gender & Age & Dep & Anx \\
\hline
Grayscale & \cellcolor{green!30}.157 & \cellcolor{red!30}-.237 & \cellcolor{green!30}.301 & \cellcolor{green!30}.258  \\
Brightness & \cellcolor{red!30}-.113 & \cellcolor{red!10}-.099 & \cellcolor{red!30}-.109 & \cellcolor{white!100}  \\
Saturation & \cellcolor{red!30}-.167 & \cellcolor{green!30}.317 & \cellcolor{red!30}-.178 & \cellcolor{red!30}-.112 \\
Pleasure & \cellcolor{green!30}.216 & \cellcolor{red!30}-.103 & \cellcolor{green!10}.093 & \cellcolor{green!10}.060 \\
Arousal & \cellcolor{red!30}-.170 & \cellcolor{green!30}.169 & \cellcolor{red!30}-.155 & \cellcolor{red!10}-.097  \\
Dominance & \cellcolor{red!30}-.154 & \cellcolor{green!10}.052 & \cellcolor{green!10}.052 & .011  \\
Hue Count & \cellcolor{red!30}-.151 & \cellcolor{green!10}.057 & \cellcolor{red!30}-.229 & \cellcolor{red!10}-.127 \\
Warm Colors & \cellcolor{green!10}.156 & \cellcolor{red!10}-.091 & \cellcolor{white!100} & \cellcolor{green!10}.083 \\
\hline
\rowcolor{Gray} \multicolumn{1}{l}{\textbf{Aesthetics}} & Gender & Age & Dep & Anx\\ 
Color Harmony & \cellcolor{white!100} & \cellcolor{green!30}.201 & \cellcolor{red!30}-.151 & \cellcolor{red!30}-.135 \\
Motion Blur & \cellcolor{red!30}-.153 & \cellcolor{green!30}.260 & \cellcolor{red!10}-.090 & \cellcolor{red!30}-.142 \\
Lighting & \cellcolor{red!10}-.062 & \cellcolor{green!30}.236 & \cellcolor{red!30}-.212 & \cellcolor{red!30}-.151 \\
Content & \cellcolor{white!100} & \cellcolor{green!30}.176 & \cellcolor{red!30}-.109 & \cellcolor{red!10}-.068 \\
Repetition & \cellcolor{red!30}-.121 & \cellcolor{green!30}.157 & \cellcolor{red!10}-.096 & \cellcolor{red!10}-.087 \\
Depth of Field & \cellcolor{red!10}-.067 & \cellcolor{green!30}.154 & \cellcolor{red!30}-.102 & \cellcolor{red!10}-.061 \\
Vivid Color & \cellcolor{red!30}-.162 & \cellcolor{green!30}.283 & \cellcolor{red!30}-.275 & \cellcolor{red!30}-.189 \\
Symmetry & \cellcolor{green!30}.120 & \cellcolor{green!30}.282 & \cellcolor{red!30}-.274 & \cellcolor{red!30}-.189 \\
Object Emphasis & \cellcolor{green!30}.136 & \cellcolor{white!100} & \cellcolor{green!10}.093 & \cellcolor{white!100} \\
Balancing Element & \cellcolor{green!10}.039 & \cellcolor{green!10}.089 & \cellcolor{white!100} & \cellcolor{white!100} \\
Overall Aesth. Score & \cellcolor{red!10}-.067 & \cellcolor{green!30}.239 & \cellcolor{red!30}-.155 & \cellcolor{red!30}-.124 \\
\hline
\rowcolor{Gray} \multicolumn{1}{l}{\textbf{Meta}} & Gender & Age & Dep & Anx   \\
\hline
\% Image posts & \cellcolor{green!10}.049 & \cellcolor{red!30}-.136 & \cellcolor{green!30}.128 & \cellcolor{green!30}.276 \\
\% Posts with People & \cellcolor{white!100} & \cellcolor{red!30}-.195 & \cellcolor{white!100} & \cellcolor{green!10}.066 \\

\hline
\end{tabular}
\caption{\label{t:postcorr} Posted Images}
\end{subtable}



\begin{subtable}{\columnwidth}
\begin{tabular}{|l!{\vrule width 1pt}c|c!{\vrule width 1pt}c|c|}
\hline
\rowcolor{DGray}
\textbf{Feature} & \multicolumn{2}{c!{\vrule width 1pt}}{\textbf{Demographics}} & \multicolumn{2}{c|}{\textbf{MentalHealth}}\\
\hline
\rowcolor{Gray}
\textbf{Colors}  & Gender & Age & Dep & Anx \\
\hline
Brightness  & \cellcolor{white!100} & \cellcolor{white!100} & \cellcolor{green!10}.043 & \cellcolor{green!10}.045 \\ 
Contrast  & \cellcolor{red!10}.033 & \cellcolor{green!10}-.08 & \cellcolor{white!100} & \cellcolor{red!10}-.011 \\ 
Hue   & \cellcolor{white!100} & \cellcolor{white!100} & \cellcolor{white!100} & \cellcolor{green!10}.044 \\
\hline
\rowcolor{Gray}
\textbf{Image Composition}   & Gender & Age & Dep & Anx\\
\hline
Hue Count  & \cellcolor{white!100} & \cellcolor{white!100} & \cellcolor{white!100} & \cellcolor{green!10}.045 \\
Visual Weight   & \cellcolor{white!100} & \cellcolor{white!100} & \cellcolor{green!10}.043 & \cellcolor{green!10}.045 \\
\hline
\rowcolor{Gray}
Image Type  & Gender & Age & Dep & Anx  \\
\hline
One Face  & \cellcolor{red!30}{-.532} & \cellcolor{red!30}{-.300} & \cellcolor{green!30}.142 & \cellcolor{green!10}.077 \\
Num. of Faces  & \cellcolor{red!30}{.165} & \cellcolor{red!30}-.242 &\cellcolor{red!30}-.121 & \cellcolor{white!100} \\
\hline
\rowcolor{Gray}
\textbf{Facial Presentation}  & Gender & Age & Dep & Anx\\
\hline
Reading Glasses  & \cellcolor{green!10}{.091} & \cellcolor{red!30}{-.581} & \cellcolor{green!10}{.064} & \cellcolor{green!10}{.064} \\
Sunglasses  & \cellcolor{white!100} & \cellcolor{red!30}{-.758} & \cellcolor{green!10}{.068} & \cellcolor{green!10}{.071}  \\
Pitch Angle  & \cellcolor{green!30}{.274} & \cellcolor{red!30}{-.518} & \cellcolor{white!100} & \cellcolor{white!100} \\
Yaw Angle  & \cellcolor{white!100} & \cellcolor{white!100} & \cellcolor{white!100} & \cellcolor{green!10}{.052} \\
Face Ratio   & \cellcolor{white!100} & \cellcolor{white!100} & \cellcolor{green!30}.136 & \cellcolor{green!10}.083 \\
\hline
\rowcolor{Gray}
\textbf{Facial Expressions}   & Gender & Age & Dep & Anx \\
\hline
Smiling   & \cellcolor{white!100} & \cellcolor{white!100} & \cellcolor{red!30}{-.185} & \cellcolor{green!10}.074 \\
Anger  & \cellcolor{red!30}{-.258} & \cellcolor{red!30}.349 & \cellcolor{green!10}{.082} & \cellcolor{white!100} \\
Disgust  & \cellcolor{red!30}{-.431} & \cellcolor{red!30}{-.137} & \cellcolor{white!100} & \cellcolor{white!100} \\
Fear  & \cellcolor{red!30}{-.281} & \cellcolor{green!30}{.148} & \cellcolor{white!100} & \cellcolor{white!100} \\
Joy  & \cellcolor{green!30}{.135} & \cellcolor{red!10}{-.066} & \cellcolor{red!30}{-.149} & \cellcolor{white!100} \\
Neutral  & \cellcolor{white!100} & \cellcolor{green!30}{.237} & \cellcolor{green!30}{.149} &\cellcolor{green!10}{.064} \\
Expressiveness  & \cellcolor{white!100} & \cellcolor{red!10}{-.082} & \cellcolor{red!30}{-.143} & \cellcolor{white!100} \\
Negative Mood  & \cellcolor{red!30}{-.166} & \cellcolor{red!10}-.078 & \cellcolor{white!100} & \cellcolor{white!100} \\
Positive Mood  & \cellcolor{green!30}{.126} & \cellcolor{white!100} & \cellcolor{red!30}{-.136} & \cellcolor{white!100} \\
\hline
\end{tabular}
\caption{\label{t:profcorr} Profile Images}
\end{subtable}

\caption{\label{t:twttextcor} Pearson correlations between color and aesthetic features extracted from posted and profile images and mental health conditions, and with age and gender (coded as 1 for female, 0 for male) separately and used age as a continuous variable. Correlations for mental health conditions are controlled for age, gender and other mental health condition. Positive correlations are highlighted with green ($p<.01$, two-tailed t-test) and negative correlations with red ($p<.01$, two-tailed t-test). Correlations which are not significant are not presented. }

\end{table}

\begin{table*}[tp!]
\small
\centering
\resizebox{\textwidth}{!}{
\begin{tabular}{|>{\arraybackslash}m{0.1\columnwidth}|>{\arraybackslash}m{0.1\columnwidth}|>{\arraybackslash}m{0.8\columnwidth}||>{\arraybackslash}m{0.1\columnwidth}|>{\arraybackslash}m{0.1\columnwidth}|>{\arraybackslash}m{0.8\columnwidth}|}
\hline
\rowcolor{DGray}
\small{\textbf{$r$\_dep}} &  \small{\textbf{$r$\_anx}} & \small{\textbf{Image Tag Clusters}} & \small{\textbf{$r$\_dep}} & \small{\textbf{$r$\_anx}} & \small{\textbf{Image Tag Clusters}} \\
\hline 
\rowcolor{Gray} \multicolumn{3}{l}{\small{\textbf{(More depressed and anxious)}}} & \multicolumn{3}{l}{\small{\textbf{(Less depressed and anxious)}}} \\
.196 & .163 & 3d, alphabet, book, capital, document, font, pen, text, typescript & .243 & .151 & action, active, aerobics, athlete, balance\_beam, ball, ballplayer, barbell, baseball \\
\rowcolor{LGray} .120 & .124 & affenpinscher, american\_staffordshire\_terrier, and, appenzeller, arctic\_fox, australian\_terrier & .233 & .129 & audience, championship, cheering, competition, crowd, event, flag, lights, match \\
---- & .122 & adult, attractive, beard, brunette, caucasian, face, fashion, glamour, gorgeous & .057 & .114 & artistic, astrology, astronomy, atmosphere, azure, buildings, business\_district \\
\rowcolor{LGray} .147 &  .109 & animals, cat, domestic\_cat, egyptian\_cat, eye, feline, fur & .134 & .100 & apartment, bed, bedclothes, bedroom, bedroom\_furniture, blanket  \\ 
 .141 & .096 & angora, badger, bunny, cavy, easter, footed\_ferret, fox\_squirrel, guinea\_pig & .096 & .083 & asphalt, avenue, broom, cleaning\_implement, expressway, highway \\
\rowcolor{LGray} ---- & .062 & businessman, businesspeople, confident, corporate, executive, handsome, manager & ---- & .076 & button, buttons, circle, design, glossy, graphic, icon, icons, internet \\
 ---- & .059 & baby, boy, boyfriend, brother, buddy, child, childhood, children, couple & .053 & .073 & alp, alpine, alps, autumn, canyon, cascade, cold, creek, crystal \\
\rowcolor{LGray} .110 & .057 & african\_chameleon, african\_crocodile, agama, agamid, alligator, american\_alligator & .079 & .064 & aroma, caffeine, cappuccino, china, coffee, coffee\_mug, crockery, cup \\
.063 & ---- & bank, banking, bill, book\_jacket, cash, comic\_book, currency, finance, financial & .057 & .056 & cuisine, delicious, dinner, dish, gourmet, lunch, meal, meat, plate, restaurant, tasty, vegetable \\
\rowcolor{LGray} .052 & ---- & aquatic\_mammal, dugong, eared\_seal, electric\_ray, great\_white\_shark, hammerhead, ray & .064 & .052 & automobile, car, convertible, drive, motor, motor\_vehicle, speed, sports\_car \\
.050 & ---- & ape, capuchin, chimpanzee, gorilla, macaque, marmoset, monkey, orangutan, primate & .054 & .048 & bicycle, bike, built, carriage, cart, horse\_cart, mountain\_bike, minibike, moped \\
\rowcolor{LGray} & & & .052 & .051 & bagel, baked, baked\_goods, bakery, bread,  breakfast, bun, burger, cheeseburger, cracker \\
& & & .061 & ---- & bay, beach, coast, coastline, fence, idyllic, island, obstruction, ocean, palm, paradise, relaxation, resort \\\hline
\end{tabular}
}
\caption{Pearson correlations between Imagga tag clusters extracted from posted pictures and mental health conditions. All correlations are significant at $p<.05$, two-tailed t-test, Benjamini-Hochberg corrected. Results for depression and anxiety are controlled for age and gender. Tags are sorted by occurrence in our data set within a cluster. `--` indicates that the tag cluster is not significantly associated with Depression (dep) and/or Anxiety (anx).}
\label{t:posttopics}
\end{table*}

\section{Analysis}

We perform univariate correlation tests between each feature and mental health condition to uncover the associated image features. We control for age and gender using partial correlation so that our analysis is not skewed by any potential demographic bias in the data. Additionally, in content analysis, as depression and anxiety are highly inter-correlated ($r$=.67), we control for the other trait in order to isolate the unique attributes of each dimension. We adjust for multiple comparisons using Benjamini-Hochberg multi-test correction.

Results for our analysis on posted images using image colors and aesthetic features on TwitterText data set are presented in Tables~\ref{t:postcorr}, using 50 Imagga topics in Table~\ref{t:posttopics}, and analysis on profile images are presented in~\ref{t:profcorr}. The same set of experiments on mental health outcomes was conducted on the TwitterSurvey data set, but were no longer significant when controlling for multiple comparisons. This shows the need for these behaviors to be studied using larger samples. However, we validated the language prediction model to predict scores on TwitterSurvey, which yielded a correlation of .12 and .20 with depression and anxiety respectively. Compared to other psychological outcomes such as personality, these correlations are consistent e.g.  \cite{segalin2017your,reece2017instagram,jaidka2018facebook}.

\subsection{Posted Images}

In Figure \ref{fig:dep_post_eg}, we visualize random images from users with top and bottom 25\% percentile of both depression and anxiety scores. Users scoring high in both depression and anxiety have a preference to post images which are grayscale, less intense in colors (low saturation) and in the diversity of colors (low hue count), low in arousal and low on all attributes of aesthetics. Further, users high in depression post images which emphasize foreground objects and are low in brightness.  Content analysis shows that users high in both traits are posting images containing text ($3d$, $alphabet$..), images of animals ($animals$, $cat$..), while users low in depression and anxiety post images of sports ($action$,$audience$..), nature ($artistic$, $astrology$), every day things ($apartment$, $bed$..), meals ($cuisine$, $delicious$..), motor vehicles ($automobile$, $car$..), outdoor activities ($bicycle$, $bike$..). We note that the associations are stronger for most topics for depression, indicating this outcome is more strongly associated with image posting preferences.

Apart from these themes, anxious users specifically are seen to post portraits and selfies (tagged as $adult$, $attractive$..), corporate environments ($businessman$, $corporate$..), and depressed users are seen to post more animal related images ($ape$, $aquatic\_mammal$..). Whereas people low in depression score post vacation related images ($beach$,$coast$..). 

\subsection{Profile Images}
Choice of profile pictures uncovers more insight into the behavior of users with mental health conditions; while depressed users preferred images which are not sharp and which do not show face, anxious users usually chose sharper images with multiple people in them. 

\section{Prediction}

Finally, we investigate the accuracy of visual features in predicting the mental health conditions. We use linear regression with ElasticNet regularization  \cite{zou05} as our prediction algorithm and report results on 10 fold cross-validation (sampled such that users in the training set are not present in the test set). Performance is measured using Pearson correlation (and Mean Squared Error in brackets) across the 10 folds. 

We also explore whether incorporating author attributes such as age and gender in the joint modeling of depression and anxiety can improve prediction performance. We achieve this by introducing a different set of regularization constraints in the ElasticNet optimization function, a method known as $l2/1$ norm regularization  \cite{liu2009multi} (denoted as MT in the results). In single-task learning, we used image features as predictors and each mental health variable as outcome in a separate regression with ElasticNet regularization. For multi-task learning, we used both demographics (age and gender) and mental health outcomes (depression or anxiety) in the same regression using L1/L2 mixed-norm as regularizer to exploit the underlying implicit correlations in mental health and demographic factors. The performance on the TwitterText data set is shown in Table \ref{t:twttext} and on the TwitterSurvey data set is shown in Table \ref{t:twtsur}. For single-task (ST), we tried support vector regression and L1 and L2 regularization for linear regression and found no significant difference in results. To be consistent in comparing the same methods in ST and MT, we used ElasticNet regularization and linear regression for both. Combination of several feature sets is also tested and termed as Combination in the Tables. 

We find that multi-task learning offers significant gains over single task learning for posted images (see Table \ref{t:textpostpred}). While VGG penultimate features do not perform as well as others at predicting depression, multi-task learning boosts the performance by almost 80\%. MT shows an advantage for both depression ($r$ = .619 for MT vs .554 for ST) and anxiety ($r$ = .580 vs .532) in Table \ref{t:textpostpred}. While colors and aesthetics do not see a big jump from single-task to multi-task learning, deep learning based methods see a drastic improvement. However, this is not the case with profile images; multi-task learning shows very similar performance as single-task learning. 

On profile images, image-based demographic predictions outperform other features, due to the dependence of mental health conditions on age, gender and race, followed by facial expressions and the aesthetic attributed of the profile image. Predicting anxiety using profile images is more accurate than predicting depression, suggesting that a single profile image might be able to provide more information about anxiety. However, a set of posted images are required to predict depression with a reasonable accuracy.

We then examine if using text-predicted outcomes as proxies for survey labels can improve performance in predicting more reliable survey based outcomes (not shown in any Table). We evaluated the performance of models trained on TwitterText when tested on TwitterSurvey, and found that the performance ($r = .164$ using aesthetics) is similar to the best performing model trained and tested on TwitterSurvey ($r=.167$) for depression, but however outperforms the corresponding model for anxiety ($r=.223$ when transfer learning from TwitterText to TwitterSurvey vs. $r=.175$ when trained and tested on TwitterSurvey). This shows that text-predicted labels can be used as proxies in analyzing image-posting behavior when studying mental health. 

\begin{table}[t!]
\begin{subtable}{\columnwidth}
\centering
\resizebox{\columnwidth}{!}{
\begin{tabular}{|c|c|cc|cc|}
\hline
\multirow{2}{*}{Feature set} & \multirow{2}{*}{\# Feat} & \multicolumn{2}{c|}{Depression} & \multicolumn{2}{c|}{Anxiety} \\ \cline{3-6} 
                             &                          & ST              & MT            & ST            & MT           \\ \hline
Colors                       & 44                       & .446 (.811)     & .449 (.802)   & .441 (.815)   & .446 (.803)  \\
Aesthetics                   & 10                       & .434 (.818)     & .434 (.810)   & .380 (.866)   & .377 (.857)  \\
Imagga                       & 500                        & .443 (.836)     & .509 (.742)   & .426 (.837)   & .483 (.766)  \\
VGG Penultimate (fc7)        & 4096                        & .343 (1.022)    & .603 (.644)   & .351 (.890)   & .555 (.693)  \\
VGG Classes                  & 1000                        & .438 (.821)     & .520 (.731)   & .442 (.811)   & .505 (.747)  \\ \hline
Combination                  & 5                        & .554 (.689)     & .619 (.613)   & .532 (.715)   & .580 (.661)  \\ \hline
\end{tabular}
}
\caption{ \label{t:textpostpred} Posted Images}
\end{subtable}

\begin{subtable}{\columnwidth}
\centering
\resizebox{\columnwidth}{!}{
\begin{tabular}{|c|c|cc|cc|}
\hline
\multirow{2}{*}{Feature set}      & \multirow{2}{*}{\# Feat} & \multicolumn{2}{c|}{Depression}               & \multicolumn{2}{c|}{Anxiety}                   \\ \cline{3-6} 
                                  &                          & \multicolumn{1}{c}{ST}          & MT          & \multicolumn{1}{c}{ST}          & MT           \\ \hline
Colors                            & 44                       & .084 (.997)                     & .101 (.989) & .133 (.986)                     & .146 (.978)  \\
Image Composition                 & 10                       & .038 (1.006)                    & .046 (.997) & .056 (1.001)                    & .059 (.996)  \\
Image Type                        & 5                        & .069 (1.001)                    & .070 (.995) & .046 (1.002)                    & .049 (.997)  \\
Image Demographics                & 5                        & .254 (.939)                     & .255 (.922) & .403 (.839)                     & .403 (.846)  \\
Facial Presentation               & 7                        & .056 (1.001)                    & .056 (.984) & .059 (1.000)                    & .061 (1.007) \\
Facial Expressions                & 14                       & .170 (.976)                     & .174 (.938) & .133 (.989)                     & .137 (.979)  \\
Aesthetics                        & 12                       & .111 (.993)                     & .114 (.986) & .100 (.994)                     & .108 (.988)  \\ \hline
\multicolumn{1}{|c|}{Combination} & \multicolumn{1}{c|}{7}   & \multicolumn{1}{c}{.305 (.880)} & .311 (.873) & \multicolumn{1}{c}{.427 (.815)} & .429 (.813)  \\ \hline
\end{tabular}
}
\caption{ \label{t:textprofpred} Profile Images}
\end{subtable}

\caption{\label{t:twttext} \textbf{TwitterText:} Prediction results for mental health conditions with all features. Performance is measured using Pearson correlation (and MSE in parenthesis) in 10-fold cross-validation.  \textbf{ST} represents Single-Task learning and \textbf{MT} represents Multi-Task learning}

\end{table}

\begin{table}[t!]
		\centering
		\resizebox{\columnwidth}{!}{
			\begin{tabular}{|c|c|cc|cc|}
\hline
\multirow{2}{*}{Feature set} & \multirow{2}{*}{\# Feat} & \multicolumn{2}{c|}{Depression} & \multicolumn{2}{c|}{Anxiety} \\ \cline{3-6} 
                             &                          & ST              & MT            & ST            & MT           \\ \hline
Colors                       & 44                       & .070 (1.06)   & .113 (.996)  & .120 (1.03)     & .086 (1.02)   \\
Aesthetics                   & 10                       & .065 (1.07)   & .068 (1.01)  & .038 (1.00)     & .130 (.998)   \\
Imagga                       & 500                        & .157 (1.06)   & .160 (1.00)  & .104 (1.23)     & .128 (1.01)   \\
VGG Penultimate (fc7)        & 4096                        & .062  (1.06)   & .143 (.999)  & .118 (1.16)    & .126 (1.02)   \\
VGG Classes                  & 1000                          & .036 (1.11)   & .045 (.999)  & .140 (1.01)     & .148 (1.00) \\ \hline
Combination                  & 5                          & .151 (1.01)   & .167 (.975)  & .167 (1.01)     & .175 (1.00) \\ \hline
\end{tabular}
		}

\caption{ \textbf{TwitterSurvey:} Prediction results for mental health conditions with all features on posted images. Performance is measured using Pearson correlation (and MSE in parenthesis) in 10-fold cross-validation.  \textbf{ST} represents Single-Task learning and \textbf{MT} represents Multi-Task learning}
\label{t:twtsur}
\end{table}

\section{Discussion}

In this paper, we explored how depression and anxiety traits can be automatically inferred by just looking at images that users post and set as profile pictures. We compared five different visual feature sets (extracted from posted images and profile pictures) and the findings about image features associated with mental illness in large part confirm previous findings about the manifestations about depression and anxiety. 

Regarding depression, perhaps with the highest face validity, in profile images of depressed users the facial expressions show fewer signs of positive mood (less joy and smiling), and are appear more neutral and less expressive. Previous findings suggest that social desirability and self-presentation biases generally discourage the sharing of negative information about the self (such as negative emotion in a society that values positivity), instead, negative emotion is often manifested as a lack of expression of positive emotion  \cite{dibble2013sharing}, as evident in this study. Additionally, depressed individuals’ profile pictures are marked by the fact that they are more likely to contain a single face (that of the user), rather than show the user surrounded by friends. In fact, focus on the self is one of the most consistent markers of negative emotionality and depression in particular  \cite{tackman2018depression}; in language studies of depression, the use of first personal singular pronouns has emerged as one of the most dependable markers of depression  \cite{holtzman2017meta} – the single user profile picture may very well be its analogue in the world of image posts.

In posted images, depressed individuals post pictures that are generally less aesthetically pleasing, extending across features such as color harmony,  lighting, content, repetition, depth of field, vividness of color and symmetry in the pictures. These findings suggest perceptual symptoms of depression, leading to picture choices that lack coherence and traditional markers of beauty -- likely because such lack of order is experienced to be more  representative of the psychological reality of depression, which is often without orientation, order and appreciation of beauty. Finally, the nominally highest correlation among all image features associated with depression is observed in the increased use of grayscale images, which may again be the visual expression of the undifferentiated, monotone emotional experience that marks depression. The automatically extracted image features that track the emotional circumplex (pleasure and arousal) fail to fully capture the depressed experience, as depressed posted images show less “arousal” (very much as expected in depression), but nominally more “pleasure.”

In terms of Imagga tag clusters of posted images, depressed users choose images with more text content and animals, neither of which is an obvious marker of depression. Again depression is more strongly manifested in what posted image content does $not$ cover, which includes a variety of markers of the psychologically well-adjusted life: sports (specifically team sports), recreational activities that include outdoor activities, driving, bicycles and beaches and vacation, in addition few references to food and beverages. 

Regarding anxiety, the facial expression in profile picture do not show fewer positive emotions, in fact, nominally more smiling is observed. Frequently, anxiety includes elements of social anxiety, which may interact with the social desirability biases inherent in social media sharing to generate a facade of well-being – anxious people are frequently referred to as ``the worried well;'' their profile images show a much less clearer set of associations that could distinguish them from the norm (all correlations $< .1$). Similar to depressed users, anxious users’ images are more grayscale and lack aesthetic cohesion across a variety of image features – here too suggesting a lack of responsiveness to traditional markers of beauty and emotional balance. Again, the emotional circumplex image features (pleasure and arousal) fail to fully capture the anxious experience, as anxious individuals are generally understood to experience more (not less) “arousal,” and less (not more) “pleasure.”

In terms of Imagga tag clusters of posted images, the profile in large part overlaps with that of depression with generally lower effect sizes compared to normative users, suggesting that anxiety is not as observable in the image choices as depression. Anxious users additionally seem to post marginally more content related to family and work; aspects of social anxiety may here manifest as the wish to signal fulfillment of typical role obligations.

We observe that multi-task learning improves the performance of mental health assessment confirming the need to include the personal characteristics of users while building automatic models  \cite{degens2017see}. Even though profile images do not offer high performance in terms of predicting mental health conditions, they offer insights as described above. Combined with the findings from posted images, such cues could be used in improving the automatic assessment of mental health conditions. 

A recent meta-analysis of predicting Big-Five personality from digital traces \cite{settanni2018predicting} shows that the Big-Five personality dimensions can be predicted from social media traces at accuracies ranging from r = .29 to .40 (generally in cross-validation frameworks on the same data), which roughly match the highest correlations that are generally found between psychological traits and observable behaviors in the large psychological literature (r \~ .3). Depression is more nuanced and a less broad characteristic than any of the Big-Five personality traits, therefore, observing a cross-validated prediction performance of r = .32 suggests very decent model performance. When the same Facebook model is applied to the TwitterSurvey dataset, the observed correlations with the TwitterSurvey survey estimates are lower  (r = .12 and .20 for depression and anxiety, respectively). This is likely due to the fact that Twitter and Facebook have differences in language, both in terms of vocabulary (e.g. emoticons) and subject matter use \cite{jaidka2018facebook,zhong2017wearing,guntuku2018understanding}. Importantly, these highly significant correlations (p $<$ .01) nevertheless demonstrate that the Facebook prediction models encode significant mental health information that can be used to estimate the mental health status of Twitter users. 

\subsection{Limitations and Future Work}

Our work has some limitations: first, though we show that the weak text-labeled mental health scores on TwitterText are reliable, by testing a model trained on TwitterText on TwitterSurvey, where ground truth was collected using surveys, further work on creating large-scale datasets is required to uncover other dimensions of depressed and anxious behavior using stronger measures. Further, we use only one profile picture per user in our analysis. It would be a promising research direction to conduct a study with experience sampling where images across time are collected along with users' response to questionnaires to study the temporal orientation of mental health conditions and how their social media imagery changes as a consequence, potentially using smartphone based data collection  \cite{singh2016cooperative}. Further, we consider only the image-based features due to the initial premise, which we've seen to be true in our results, that visual imagery can uncover several strong signals for mental health conditions. Users' social network features, coupled with image posting behavior and also text, could further enhance the understanding about the manifestation of depression and anxiety  \cite{hong2017user}. 

Another direction of work could assess if predictive/automatic assessments of mental health could outperform third-party human assessments, as has been seen in other works  \cite{segalin2017your}. Also, we did not collect the date of first onset of depression from the users; this data could be used to test if social-media based machine assessments could be used as an early screening tools to help clinicians in identifying at patients at risk of mental health disorders, which usually manifest early on platforms like social media \cite{deChoudhury2013social}. Finally, we also do not consider predicting depression and anxiety as a classification task as it is not easy to binarise mental health conditions using median or quartile-split  \cite{beck1996beck} and a continuous score can often be helpful in personalizing the diagnosis and treatment for patients, and also serve as a risk assessment tool for providers.

As a benchmark, the diagnosis and differentiation of mental health conditions is difficult and costly. The gold standard of clinical psychological diagnosis used in the therapy effectiveness literature are “Structured Clinical Interviews for the Diagnostic and Statistical Manual of Mental Disorders” (SCIDs). They involve a trained clinician’s interviewing the patient in person for at least about an hour, using a guide that serves as a decision tree. Passive approaches that analyze digital traces left online are at this point substantially less powerful in their ability to detect the presence of mental health conditions, in part because they are unobtrusive (unlike validated clinical surveys like the PHQ-9, which have 80\%+ sensitivity and 90\%+ specificity compared to the SCID baselines  \cite{gilbody2007screening}). In terms of improving the unobtrusive detection of mental health conditions, the next step for the field appears to be multi-modal approaches that combine the relative power of different feature spaces. Specifically, text-based analyses of social media ought be combined with those of text messages and other instant messaging communication, and with analyses of image content being shared, as in this study. Importantly, the more domains of communication are covered outside of social media, the more such data feeds can cover lack of information being shared on social media, as may occur during sensor depression. In addition, phone or wearable data can give estimates as to the users activity levels and sleeping hours (which are closely related to diagnosis criteria for depression). Further, the results in this study do not imply causality, but are correlational.

Our paper studies the significance of using visual traces on social media to glean insight and develop predictive models to automatically assess users' mental health conditions. While there are several flip sides of such technologies when used for incorrect motives such as pre-screening by insurance or employers to discriminate against specific individuals, this research is very useful to develop techniques of providing summarized feedback both to social media users and their clinicians about social media activity to help them quantify the extent to which their mental health condition has shown its footprint in the digital world; we hypothesize that such a consent-based sharing can help in more productive therapy sessions, apart from the opportunity that the user has to self-regulate. Thus, data collection, processing and dissemination of resulting models has to keep the privacy of users as top priority and should be for discovering new insights into the manifestation of mental health conditions and to better assist clinicians. 

\section{Conclusion}
We analyzed image posting and profile picture preferences using interpretable Imagga tag clusters, colors. aesthetic and facial features with the aim of identifying the way and extent to which they reveal users' mental health conditions based on the images they post and select as profile pictures on Twitter. For example, images posted by depressed and anxious users tend to be dominated by grayscale, low arousal images lacking in aesthetic sense, while anxious users are characterized by posting more images compared to regular users and users with depression. The choice of profile pictures uncovers that depressed users prefer images which are not sharp and which do not contain a face, anxious users usually chose sharper images with multiple faces in them. Results indicate that multi-task learning gives significant improvements in performance for modelling mental health conditions jointly with demographics (here age and gender), factors which clinicians usually consider while diagnosing patients. Further, models trained on larger data sets using text-predicted outcomes show reliable performance when predicting more reliable survey based mental health outcomes. Our results offer new insights and a promising avenue for future mental health research of individuals.

\bibliographystyle{aaai}
\bibliography{FULL-GuntukuS.1011}

\end{document}